\documentclass[aps,prl,twocolumn,groupedaddress,showpacs]{revtex4-1} 
\usepackage{graphics}
\usepackage{graphicx}
\usepackage{amsmath}
\usepackage{amssymb}
\usepackage{amsthm}
\usepackage{epsfig}
\usepackage{dsfont}
\usepackage{xcolor} 
\usepackage[T1]{fontenc}
\usepackage{lmodern}
\begin{document}
\title{Towards room-temperature superconductivity}
\author{Jacob Szeftel$^1$}
\email[corresponding author :\quad]{jszeftel@lpqm.ens-cachan.fr}
\author{Nicolas Sandeau$^2$}
\author{Michel Abou Ghantous$^3$}
\author{Muhammad El-Saba$^4$}
\affiliation{$^1$ENS Paris-Saclay/LuMIn, 4 avenue des Sciences, 91190 Gif-sur-Yvette, France}
\affiliation{$^2$Aix Marseille Univ, CNRS, Centrale Marseille, Institut Fresnel, F-13013 Marseille, France}
\affiliation{$^3$American University of Technology, AUT Halat, Highway, Lebanon}
\affiliation{$^4$Ain-Shams University, Cairo, Egypt}
\begin{abstract}
By taking advantage of a stability criterion established recently, the critical temperature $T_c$ is reckoned with help of the microscopic parameters, characterising the normal and superconducting electrons, namely the independent-electron band structure and a \textit{repulsive} two-electron force. The emphasis is laid on the sharp $T_c$ dependence upon electron concentration and inter-electron coupling, which might offer a practical route toward higher $T_c$ values and help to understand why high-$T_c$ compounds exhibit such remarkable properties.
\end{abstract}
\pacs{74.25.Bt,74.25.Jb,74.62.Bf}
\maketitle
The BCS theory\cite{sch}, despite its impressive success, does not enable one to predict\cite{mat} superconductivity occurring in any metallic compound. Such a drawback ensues from an \textit{attractive} interaction, assumed to couple electrons together, which is not only at loggerheads with the sign of the Coulomb repulsion but in addition leads to questionable conclusions to be discussed below. Therefore this work is intended at investigating the $T_c$ dependence upon the parameters, characterising the motion of electrons correlated together through a \textit{repulsive} force, within the framework of a two-fluid picture\cite{sz8} to be recalled below.\par 
	The conduction electrons comprise bound and independent electrons, in respective temperature dependent concentration $c_s(T),c_n(T)$, such that $c_0=c_s(T)+c_n(T)$ with $c_0$ being the total concentration of conduction electrons. They are organized, respectively, as a many bound electron\cite{sz5} (MBE) state, characterised by its chemical potential $\mu(c_s)$, and a Fermi gas\cite{ash} of Fermi energy $E_F(T,c_n)$. The Helmholz free energy of independent electrons per unit volume $F_n$ and $E_F$ on the one hand, and the eigenenergy per unit volume $\mathcal{E}_s(c_s)$ of bound electrons and $\mu$ on the other hand, are related\cite{ash,lan}, respectively, by $E_F=\frac{\partial F_n}{\partial c_n}$ and $\mu=\frac{\partial \mathcal{E}_s}{\partial c_s}$. Then a stable equilibrium is conditioned\cite{sz4} by Gibbs and Duhem's law
\begin{equation}
\label{gidu}
E_F(T,c_n(T))=\mu(c_s(T))\quad ,
\end{equation}
which expresses\cite{lan} that the total free energy $F_n+\mathcal{E}_s$ is minimum provided $\frac{\partial E_F}{\partial c_n}+\frac{\partial \mu}{\partial c_s}>0$. Noteworthy is that $\frac{\partial \mu}{\partial c_s}<0$ has been shown to be a prerequisite for persistent currents\cite{sz4}, thermal equilibrium\cite{sz5}, the Josephson effect\cite{sz6} and a stable\cite{sz8} superconducting phase.  Likewise, Eq.(\ref{gidu}) reads\cite{sz5,sz4} for $T=T_c$
 \begin{equation}
\label{coo}
E_F(T_c,c_0)=\mu(c_s=0)=\varepsilon_b/2\quad , 
\end{equation} 
with $\varepsilon_b$ being the energy of a \textit{bound} electron pair\cite{sz5}. Note that Eqs.(\ref{gidu},\ref{coo}) are consistent with the superconducting transition being of second order\cite{lan}, whereas it has been shown\cite{sz5} to be of first order at $T<T_c$ ($\Rightarrow E_F(T,c_0-c_s)\neq\mu(c_s)$), if the sample is flown through by a finite current.\par
	The binding energy\cite{sz5} of the superconducting state $E_b(T<T_c)$ has been worked out as
	$$E_b(T)=\int_{T}^{T_c}\left(C_s(u)-C_n(u)\right)du\quad,$$
with $C_s(T),C_n(T)$ being the electronic specific heat of a superconductor, flown through by a vanishing current\cite{sz5} and that of a degenerate Fermi gas\cite{ash}. A stable phase ($\Rightarrow E_b>0$) requires $C_s(T_c)>C_n(T_c)$, which can be secured\cite{sz8} \textit{only} by fulfilling the following condition
\begin{equation}
\label{cri}
\frac{\partial E_F}{\partial c_n}(T_c,c_0)=-\frac{\partial\mu}{\partial c_s}(0),\quad\rho'(E_F(T_c,c_0))>0\quad,
\end{equation}
with $\rho\left(\epsilon\right),\epsilon$ being the independent electron density of states and one-electron energy, respectively, and $\rho'=\frac{d\rho}{d\epsilon}$.\par
	Since the remaining analysis relies heavily on Eqs.(\ref{coo},\ref{cri}), explicit expressions are needed for $E_F(T_c,c_0),\frac{\partial E_F}{\partial c_n}(T_c,c_0),\varepsilon_b,\frac{\partial\mu}{\partial c_s}(0)$. Because the independent electrons make up a degenerate Fermi gas ($\Rightarrow T<<E_F/k_B$ with $k_B$ being Boltzmann's constant), applying the Sommerfeld expansion\cite{ash} up to $T^2$ yields
\begin{equation}
\label{somm}
\begin{array}{c} 
E_F(T_c,c_0)=E_F(0,c_0)-\frac{\rho'}{\rho}\frac{\left(\pi k_BT_c\right)^2}{6}\\	 
\frac{\partial E_F}{\partial c_n}(T_c,c_0)=\left(\rho+\rho''\frac{\left(\pi k_BT_c\right)^2}{6}\right)^{-1}
\end{array}\quad ,
\end{equation}
with $\rho=\rho(E_F(0,c_0)),\rho'=\frac{d\rho}{dE_F}(E_F(0,c_0)),\rho''=\frac{d^2\rho}{dE_F^2}(E_F(0,c_0))$. As for $\varepsilon_b,\frac{\partial\mu}{\partial c_s}(0)$, a truncated Hubbard Hamiltonian $H_K$, introduced previously\cite{ja1,ja2,ja3}, will be used. The main features of the calculation\cite{sz5} are summarised below for self-containedness.\par
	The independent electron motion is described by the Hamiltonian $H_d$
$$H_d=\sum_{k,\sigma}\epsilon(k)c^+_{k,\sigma}c_{k,\sigma}\quad .$$ 
$\epsilon(k),k$ are the one-electron energy ($\epsilon(k)=\epsilon(-k)$) and a vector of the Brillouin zone, respectively, $\sigma=\pm$ is the electron spin and the sum over $k$ is to be carried out over the whole Brillouin zone. Then $c^+_{k,\sigma},c_{k,\sigma}$ are creation and  annihilation operators on the Bloch state $\left|k,\sigma\right\rangle$ 
	$$\left|k,\sigma\right\rangle=c^+_{k,\sigma}\left|0\right\rangle\quad ,\quad \left|0\right\rangle=c_{k,\sigma}\left|k,\sigma\right\rangle\quad ,$$
with $\left|0\right\rangle$ being the no electron state. The Hamiltonian $H_K$ reads then
$$H_K=H_d+\frac{U}{N}\sum_{k,k'}c^+_{k,+}c^+_{K-k,-}c_{K-k',-}c_{k',+}\quad ,$$ 
with $N>>1,U>0$ being the number of atomic sites, making up the three-dimensional crystal, and the Hubbard constant, respectively. Note that the Hamiltonian used by Cooper\cite{cooper} is identical to $H_{K=0}$, but with $U<0$.\par
	$H_K$ sustains\cite{sz5} a single \textit{bound} pair eigenstate, the energy $\varepsilon_b(K)$ of which is obtained by solving 
\begin{equation} 
\label{coo2}
\frac{1}{U}=\frac{1}{N}\sum_{k}\frac{1}{\varepsilon_b(K)-\varepsilon(K,k)}=\int_{-t_K}^{t_K}\frac{\rho_K(\varepsilon)}{\varepsilon_b(K)-\varepsilon}d\varepsilon.  
\end{equation} 
$\pm t_K$ are the upper and lower bounds of the two-electron band, i.e. the maximum and minimum of $\varepsilon(K,k)=\epsilon(k)+\epsilon(K-k)$ over $k$, whereas $\rho_K(\varepsilon)$ is the corresponding two-electron density of states, taken equal to 
	$$\rho_K(\varepsilon)=\frac{2}{\pi t_K}\sqrt{1-\left(\frac{\varepsilon}{t_K}\right)^2}\quad .$$\par
\begin{figure}
\label{coop1}
\includegraphics[width=7 cm]{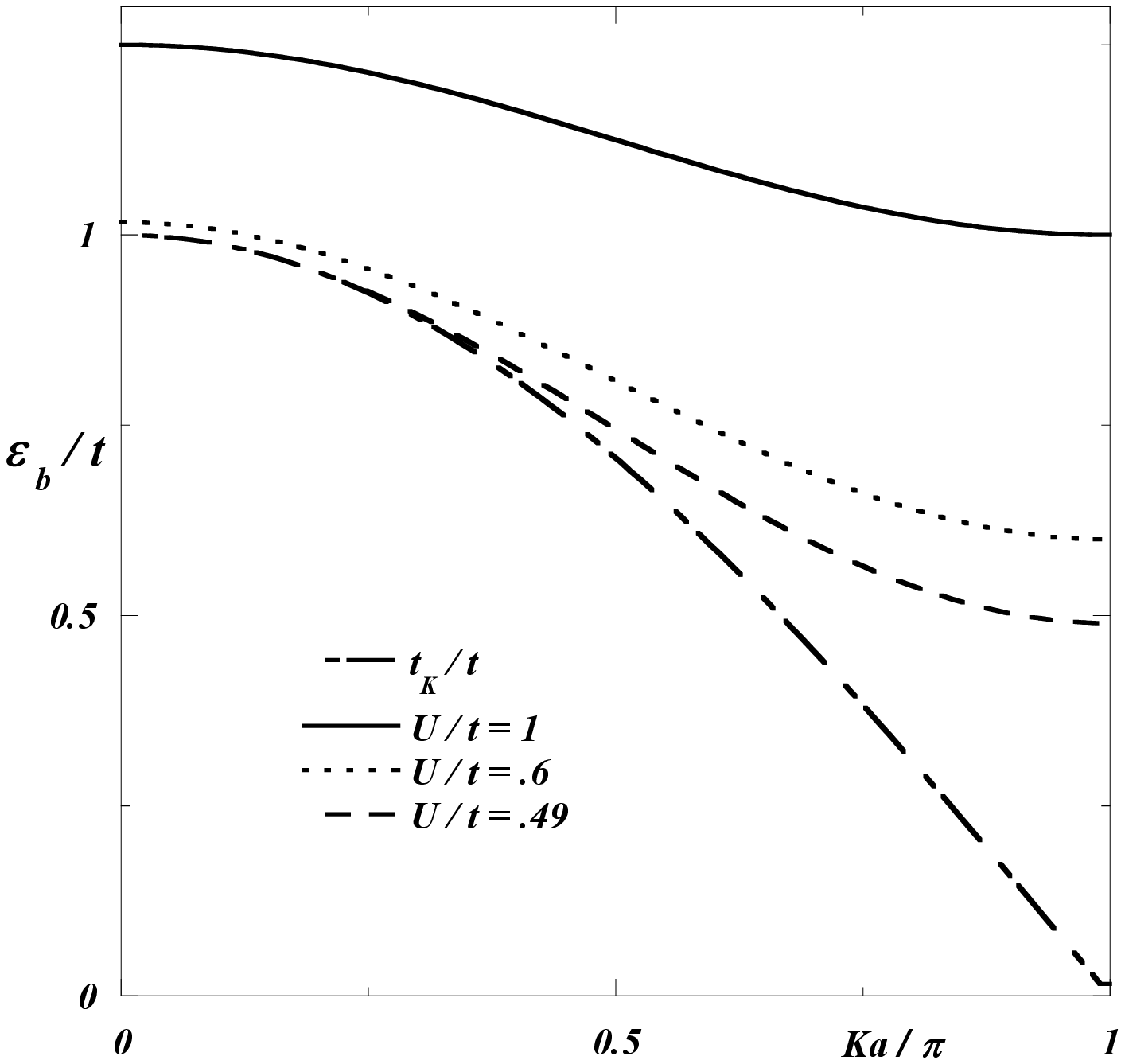}
\caption{Dispersion curves of $t_K$ as a dashed-dotted line and of $\varepsilon_b(K)$ as solid, dashed and dotted lines, associated with various $U$ values, respectively;  those data have been obtained with $t_K=t\cos\left(Ka/2\right)$, where $t,a$ are the one-electron bandwidth and the lattice parameter, respectively.}
\end{figure}

	The dispersion curves $\varepsilon_b(K)$ are plotted in Fig.1\ref{coop1}. Though Eq.(\ref{coo2}) is identical to the equation yielding the Cooper pair energy\cite{cooper}, their respective properties are quite different :
\begin{itemize}
	\item
the data in Fig.1\ref{coop1} have been calculated with $U>0$, rather than $U<0$ favoured by Cooper\cite{cooper} and BCS\cite{sch}, because, due to the inequality\cite{sz5} $U\frac{\partial\mu}{\partial c_s}<0$, choosing $U<0$ entails $\frac{\partial\mu}{\partial c_s}>0$, which has been shown \textit{not} to be consistent with persistent currents\cite{sz4}, thermal equilibrium\cite{sz5}, the Josephson effect\cite{sz6} and occurence\cite{sz8} of superconductivity. As a further consequence of $U>0$, $\varepsilon_b(K)$ shows up in the upper gap of the two-electron band structure ($\Rightarrow\varepsilon_b(K)>t_K$) rather than in the lower gap ($\Rightarrow\varepsilon_b(K=0)<-t_{K}$) in case of the Cooper pair\cite{cooper}. Nevertheless the bound pair is thermodynamically stable, because every one-electron state of energy $<E_F(T_c,c_0)$, is actually occupied, so that, due to Pauli's principle, a bound electron pair of energy $\varepsilon_b(K)=2E_F(T_c,c_0)$, according to Eq.(\ref{coo}), cannot decay into two one-electron states $\epsilon(k)<E_F,\epsilon(K-k)<E_F$;  
	\item 
	a remarkable feature in Fig.1\ref{coop1} is that $\varepsilon_b(K)\rightarrow t_K$ for $U\rightarrow t_K/2$, so that there is \textit{no} bound pair for $U<t_K/2$ (accordingly, the dashed curve is no longer defined in Fig.1\ref{coop1} for $\frac{Ka}{\pi}<.13$), in marked contrast with the opposite conclusion drawn by Cooper\cite{cooper}, that there is a Cooper pair, even for $U\rightarrow 0$. This discrepancy results from the three-dimensional Van Hove singularities, showing up at both two-electron band edges $\rho_K\left(\varepsilon\rightarrow\pm t_K\right)\propto \sqrt{t_K-\left|\varepsilon\right|}$, unlike the two-electron density of states, used by Cooper\cite{cooper} which is constant and thence displays no such singularity. Likewise the width of Cooper's two-electron band is  equal to a Debye phonon energy $2t_{K=0}=\omega_D\approx 30meV<<E_F\approx 3eV$. Hence the resulting small concentration of superconducting electrons, $\frac{c_s(T=0)}{c_0}\approx\frac{\omega_D}{E_F}\approx .01$, entails that London's length should be at least $10$ times larger than observed values\cite{sz1,sz2,sz3,sz7};
	\item
		at last Cooper's assumption $U<0$ implies $\varepsilon_b/2\neq E_F(T_c)$, which is typical of a first order transition but runs afoul at all measurements, proving conversely the superconducting transition to be of second order ($\Rightarrow \varepsilon_b/2=E_F(T_c)$ in accordance with Eq.(\ref{coo})).
\end{itemize}\par
	The bound pair of energy $\varepsilon_b(K)$ turns, at finite concentration $c_s$, into a MBE state, characterised by $\mu(c_s)$. Its properties have been calculated thanks to a variational procedure\cite{sz5}, displaying several merits with  respect to that used by BCS\cite{sch} :
\begin{itemize}
	\item 
 it shows that $\mu(0)=\varepsilon_b/2$;
	\item 
	the energy of the MBE state has been shown to be exact for $\left|U\right|\rightarrow\infty$;
	\item 
	an analytical expression has been worked out for $\frac{\partial\mu}{\partial c_s}(K,c_s=0)$ as :
\begin{equation}
\label{co2}
\frac{\partial\mu}{\partial c_s}(K,c_s=0)=-\frac{\int_{-t_K}^{t_K}\frac{\rho_K(\varepsilon)}{\left(\varepsilon_b(K)-\varepsilon\right)^3}d\varepsilon}{2\left(\int_{-t_K}^{t_K}\frac{\rho_K(\varepsilon)}{\left(\varepsilon_b(K)-\varepsilon\right)^2}d\varepsilon\right)^2}\quad .
\end{equation}
\end{itemize}\par
	The $T_c$ dependence on $c_0$ will be discussed by assigning to $\rho(\epsilon)$ the expression, valid for free electrons 
\begin{equation}
\label{free}
\rho(\epsilon)=\eta\sqrt{\epsilon-\epsilon_b}\Rightarrow c_0=\frac{2}{3}\eta\left(E_F\left(0,c_0\right)-\epsilon_b\right)^{\frac{3}{2}}\quad ,
\end{equation}
with $\eta=\frac{\sqrt{2}m^{\frac{3}{2}}V}{\pi^2\hbar^3}$, whereas $\epsilon_b,m,V=17\AA^3$ stand for the bottom of the conduction band, electron mass and volume of the unit-cell, respectively. With help of Eq.(\ref{somm}), Eqs.(\ref{coo},\ref{cri}) can be recast into a system of two equations 
\begin{equation}
\label{TcK}
\begin{array}{l}
 E_F(0,c_0)-\frac{\rho'}{\rho}\frac{\left(\pi k_BT_c\right)^2}{6}-\frac{\varepsilon_b(K)}{2}=0\\
\left(\rho+\rho''\frac{\left(\pi k_BT_c\right)^2}{6}\right)^{-1}+\frac{\partial\mu}{\partial c_s}(K,c_s=0)=0
\end{array}\quad,
\end{equation}
to be solved for the two unknowns $c_0(T_c),t_K(T_c)$ with $T_c$ being dealt with as a disposable parameter.\par
				To that end, starting values are assigned to $U,t_K$, which gives access to $\varepsilon_b(K),\frac{\partial\mu}{\partial c_s}(K,c_s=0)$) and thence to $E_F\left(0,c_0\right),\epsilon_b$ and finally to $c_0$, owing to Eqs.(\ref{coo},\ref{cri},\ref{free}). Those values of $c_0,t_K$ are then fed into Eqs.(\ref{TcK}) to launch a Newton procedure, yielding the solutions $c_0(T_c),t_K(T_c)$. The results are presented in table \ref{tab}. Since we intend to apply this analysis to high-$T_c$ compounds\cite{arm}, we have focused upon low concentrations $c_0<0.2$, which entails, in view of Eqs.(\ref{somm},\ref{free}), that $\left|\frac{\partial\mu}{\partial c_s}\right|$ takes a high value. This requires in turn $\varepsilon_b(K)\rightarrow t_K$ (see Eq.(\ref{co2})) and thence\cite{sz5} $U\rightarrow\frac{t_K}{2}$, in agreement with $\frac{t_K}{U}\approx 2$ in table \ref{tab}.\par
  A remarkable property of the data in table \ref{tab} is that $c_0,t_K$ are barely sensitive to large variations of $T_c$, i.e. $\left|\delta c_0\right|<10^{-3},\left|\delta t_K\right|<10^{-5}$ for $\delta T_c\approx 400K$. This can be understood as follows :  taking advantage of Eqs.(\ref{coo},\ref{somm},\ref{free}) results into
	$$\frac{2E_F(0,c_0)}{\varepsilon_b(K)}-1=\frac{\pi^2}{12}\left(\frac{k_BT_c}{\Delta(T_c)}\right)^2\quad ,$$
which, due to $\frac{dt_K}{dT_c}\approx 0,\Delta(T_c)\approx 1eV,T_c=400K$, yields indeed $\delta c_0=c_0(400K)-c_0(1K)\approx 10^{-3}$, in agreement with the data in table \ref{tab}. Such a result is significant in two respects, regarding high-$T_c$ compounds, for which $c_0$ can be varied over a wide range :
\begin{itemize}
	\item 
	because of $\frac{dc_0}{dT_c}\approx 0$, the one-electron band structure can be regarded safely as $c_0$ independent, which enhances the usefulness of the above analysis;
	\item 
	the large doping rate up to $\approx 0.2$ is likely to give rise to local fluctuations of $c_0$, which, in view of the utmost sensitivity of $T_c$ with respect to $c_0$, will result into a heterogeneous sample, consisting in domains, displaying $T_c$ varying from $0$ up to a few hundreds of $K$. Thus the observed $T_c$ turns out to be the upper bound of a broad distribution of $T_c$ values, associated with superconducting regions, the set of which makes up a percolation path throughout the sample. However, if the daunting challenge of making samples, wherein local $c_0$ fluctuations would be kept well below $10^{-4}$, could be overcome, this might pave the way to superconductivity at \textit{room} temperature.
\end{itemize}\par
\begin{figure}
\includegraphics[width=7 cm]{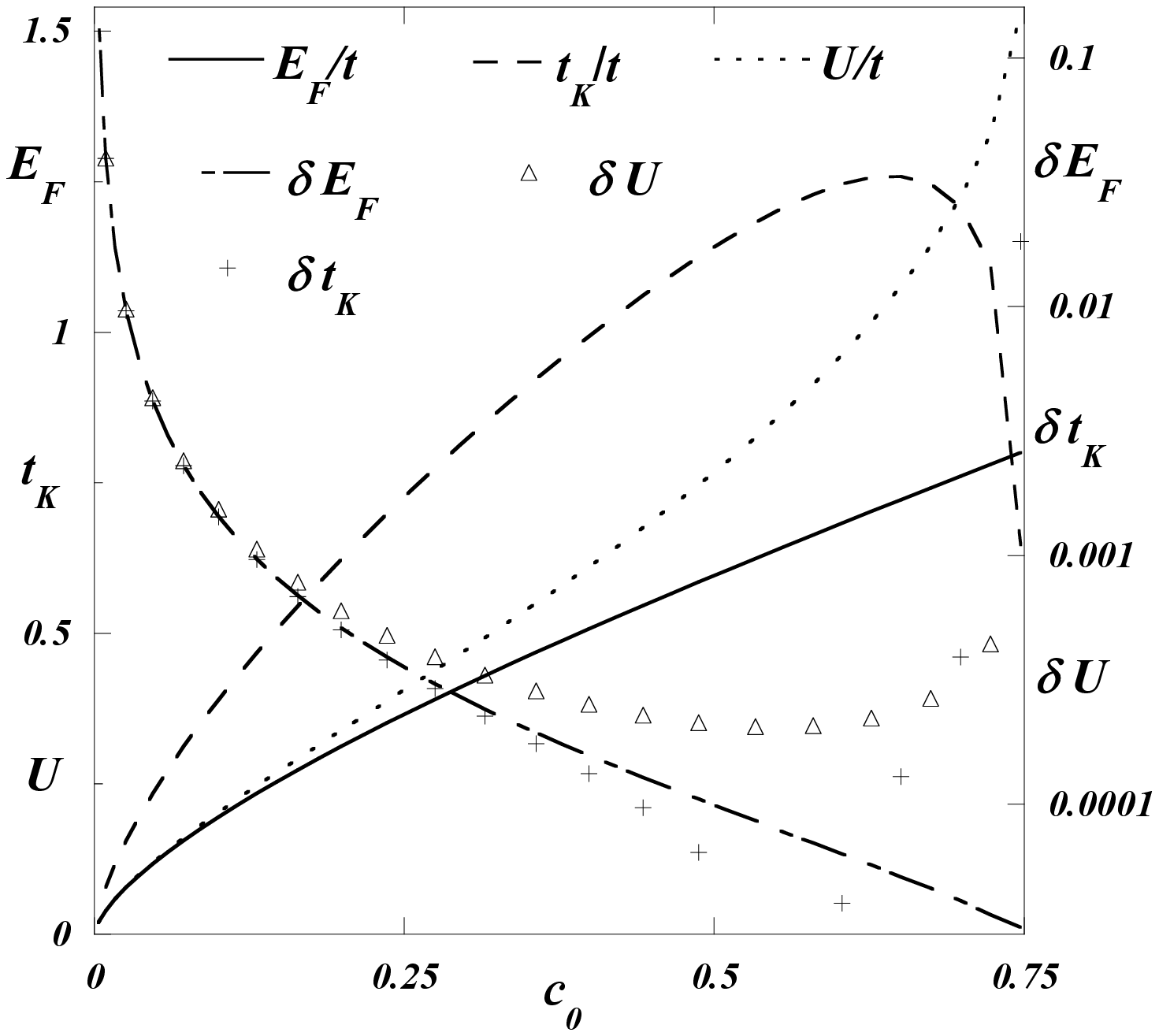}\label{uef}
\caption{Plots of $E_F(T_c,c_0),t_K(T_c,c_0),U(T_c,c_0)$ calculated for $T_c=1K$ and $t=3eV$; the unit for $c_0$ is the number of conduction electrons per atomic site; $\delta f$ with $f=E_F,t_K,U$ is defined as $\delta f=\left|1-\frac{f(300K,c_0)}{f(1K,c_0)}\right|$; the scale is linear for $E_F,t_K,U$ but logarithmic for $\delta E_F,\delta t_K,\delta U$.}
\end{figure}

\begin{table}
\caption{Solutions $c_0(T_c),t_K(T_c),\Delta(T_c)$ ($\Delta(T_c)=E_F\left(0,c_0(T_c)\right)-\epsilon_b$) of Eqs.(\ref{TcK}); $t_K,\Delta,U$ are expressed in $eV$, whereas the unit for $c_0$ is the number of conduction electrons per atomic site.}
\label{tab}
\begin{center}
\begin{tabular}{|c|c|c|c|}
\hline
 $T_c(K)$ & $c_0$ & $t_K$ & $\Delta$ \\
\hline 
 $1$ & $0.10215$  & $6$  & $1.1976$  \\
\hline
$400$ & $0.10225$ & $5.9999$ & $1.1984$  \\
\hline
\end{tabular}\quad $U=3.39$\vspace{.5cm}
\begin{tabular}{|c|c|c|c|}
\hline
 $T_c(K)$ & $c_0$ & $t_K$ & $\Delta$ \\
\hline 
 $1$ & $0.14897$  & $2$  & $1.5402$  \\
\hline
$400$ & $0.14906$ & $1.9999$ & $1.5407$  \\
\hline
\end{tabular}\quad $U=1.04$\vspace{.5cm}
\begin{tabular}{|c|c|c|c|}
\hline
 $T_c(K)$ & $c_0$ & $t_K$ & $\Delta$  \\
\hline 
 $1$ & $0.19158$  & $4$  & $1.8214$  \\
\hline
$400$ & $0.19167$ & $3.9999$ & $1.8219$  \\
\hline
\end{tabular}\quad $U=2.2$
\end{center}
\end{table}
	The $T_c$ dependence upon $U$ will be analysed with 
	$$\rho(\epsilon)=\frac{4}{\pi t}\sqrt{1-\left(1-\frac{\epsilon}{t}\right)^2}\quad ,$$ 
where $2t$ stands for the one-electron bandwidth. Our purpose is to determine the unknowns $t_K(E_F,T_c),U(E_F,T_c)$ with $E_F=E_F(T=0,c_0)$ and $c_0=\int_0^{E_F}\rho(\epsilon)d\epsilon$. To that end, Eq.(\ref{cri}) will first be solved for $t_K$ by replacing $\frac{\partial E_F}{\partial c_n}(T_c,c_0),\frac{\partial\mu}{\partial c_s}(0)$ by their expressions given by Eqs.(\ref{somm},\ref{co2}), while taking advantage of Eq.(\ref{coo}). Then the obtained $t_K$ value is fed into Eq.(\ref{coo2}) to determine $U$. The results are presented in Fig.2\ref{uef}.\par
	It can be noticed that there is no solution for $c_0>.75$, because $\frac{\partial E_F}{\partial c_n}(T_c,c_0)\approx\frac{1}{\rho}\left(E_F(0,c_0)\right)$ and $\frac{\partial\mu}{\partial c_s}(0)>\frac{U}{2}$ decrease and increase, respectively, with increasing $c_0$, so that Eq.(\ref{cri}) can no longer be fulfilled eventually. But the most significant feature is that $\delta U$ is almost insensitive to large $T_c$ variation, except for $E_F\rightarrow 0$, i.e. for $E_F$ close to the Van Hove singularity, located at the bottom of the band, which has two consequences :
\begin{itemize}
	\item 
	$c_0$ cannot be varied in most superconducting materials, apart from high-$T_c$ compounds, so that $U$ is unlikely to be equal to $U(c_0)$, indicated in Fig.2\ref{uef}. Conversely, since high-$T_c$ compounds allow for wide $c_0$ variation, $c_0$ can be tuned so that $U=U(c_0)$;
	\item
	the only possibility for a non high-$T_c$ material to turn superconducting is then offered at the bottom of the band, because $\delta U$ becomes large due to $\frac{\rho'}{\rho}(E_F\rightarrow 0)\propto \frac{1}{E_F}$ in Eq.(\ref{somm}). Such a conclusion, that superconductivity was likely to occur in the vicinity of a Van Hove singularity in low-$T_c$ materials, had already been drawn\cite{sz5} independently, based on magnetostriction data.
\end{itemize}\par
	It will be shown now that $\rho(\epsilon),\rho_K(\varepsilon)$ cannot stem from the same one-electron band. The proof is by contradiction. As a matter of fact $\rho(\epsilon)$ should read in that case
	$$\rho(\epsilon)=\frac{4}{\pi t}\sqrt{1-\left(\frac{\epsilon}{t}\right)^2}\quad .$$ 
Hence $U>0$ entails, in view of Fig.1\ref{coop1} and Eq.(\ref{coo}), that there is $\frac{\varepsilon_b}{2}=E_F>0$, which implies $\rho'(E_F)<0$ in contradiction with Eq.(\ref{cri}). Accordingly, since the two different one-electron bands, defining respectively $\rho(\epsilon),\rho_K(\varepsilon)$, display a sizeable overlap, they should in addition belong to different symmetry classes of the crystal point group, so that superconductivity cannot be observed if there are only $s$-like electrons at $E_F$ or if the point group reduces to identity. Noteworthy is that those conclusions had already been drawn empirically\cite{mat}.\par
	The critical temperature $T_c$ has been calculated for conduction electrons, coupled via a repulsive force, within a model based on conditions, expressed in Eqs.(\ref{coo},\ref{cri}). Superconductivity occurring in conventional materials has been shown to require $E_F(T_c)$ being located near a Van Hove singularity, whereas a practical route towards still higher $T_c$ values has been delineated in high-$T_c$ compounds, provided the upper bound of local $c_0$ fluctuations can be kept very low. The thermodynamical criterions in Eqs.(\ref{coo},\ref{cri}) unveil the close interplay between independent and bound electrons in giving rise to superconductivity. At last, it should be noted that Eqs.(\ref{coo},\ref{cri}) could be applied as well to any second order transition, involving only conduction electrons, such as ferromagnetism or antiferromagnetism.  

\end{document}